\documentclass[11pt,envcountsame,envcountsect]{llncs}
\usepackage[latin1]{inputenc}
\usepackage{amsmath,amsfonts,amssymb}
\usepackage[letterpaper,hmargin=1.5in,vmargin=1.75in]{geometry}
\usepackage{array}
\usepackage{here}
\usepackage{graphics}
\usepackage{multirow,graphicx,epsfig}

\begin{document}

\title{Operand Folding Hardware Multipliers}

\author{Byungchun Chung\inst{1} \and Sandra Marcello$^\star$\inst{2} \and Amir-Pasha Mirbaha\inst{3}\\
and David Naccache\inst{4} \and Karim Sabeg\inst{5}}
\titlerunning{Operand Folding Hardware Multipliers}
\authorrunning{Chung {\sl et alii.}}

\institute{%
Korea Advanced Institute of Science and Technology\\
%Department of Electrical Engineering and Computer Science\\
%Daejeon 305-701, Republic of Korea\\
\email{bcchung@nslab.kaist.ac.kr}\\
\and
{\sc thales}\\
%160, bld de Valmy BP 82\\
%92704 Colombes {\sc cedex}, France\\
\email{samarcello@hotmail.com}\\
\and
Centre micro\'electronique de Provence G. Charpak\\
%D\'epartement {\sc sas}\\
%80 Avenue de Mimet, {\sc f}-13120 Gardanne, France\\
\email{mirbaha@emse.fr}\\
\and
\'Ecole normale sup\'erieure ({\sc ens/cnrs/inria})\\
%, \'Equipe de cryptographie\\
%45 rue d'Ulm, {\sc f}-75230 Paris {\sc cedex 05}, France \\
\email{david.naccache@ens.fr}\\
\and
Universit\'e Paris 6 -- Pierre et Marie Curie\\
\email{km\_sabeg@hotmail.fr}\\
}
\maketitle

\makeatletter
\let\@thefnmark\relax
\@footnotetext{$\star$ most of the work has been done while this author was working at
the Max-Planck Institut f\H{u}r Mathematik ({\sc mpim} Bonn, Germany)}
\makeatother

\begin{abstract}

This paper describes a new accumulate-and-add multiplication algorithm. The method partitions one of the operands and re-combines the results of computations done with each of the partitions. The resulting design turns-out to be both compact and fast.\smallskip

When the operands' bit-length $m$ is 1024, the new algorithm requires only $0.194m+56$ additions (on average), this is about half the number of additions required by the classical accumulate-and-add multiplication algorithm ($\frac{m}2$).

\end{abstract}

\section{Introduction}
\label{Sec1}

Binary multiplication is one of the most fundamental operations in digital electronics. Multiplication complexity is usually measured by bit additions, assumed to have a unitary cost.\smallskip

Consider the task of multiplying two $m$-bit numbers $A$ and $B$ by repeated accumulations and additions. If $A$ and $B$ are chosen randomly ({\sl i.e.} of expected Hamming weight $w$ = $m/2$) their classical multiplication is expected to require $w(B) = m/2$ additions of $A$.\smallskip

The goal of this work is to decrease this work-factor by splitting $B$ and batch-processing its parts. The proposed algorithm is similar in spirit to
common-multiplicand multiplication ({\sc cmm}) techniques \cite{YL93}, \cite{WC95}, \cite{Yen97}, \cite{KJ94}.

\section{Proposed Multiplication Strategy}

We first extend the exponent-folding technique \cite{LC96}, suggested for exponentiation, to multiplication. A similar approach has been tried in \cite{Yen97} to fold the multiplier into halves. Here we provide an efficient and generalized operand decomposition technique, consisting in a
memory-efficient multiplier partitioning method and a fast combination method. For the sake of clarity, let us illustrate the method with a toy example. As the multiplicand $A$ is irrelevant in estimating the work-factor ($A$ only contributes a multiplicative constant), $A$ will be omitted.

\subsection{A Toy Example}

Let $m = 2 \cdot n$ and $B = {101010100011}_2 = B_2 || B_1 = b_5^2 b_4^2 b_3^2 b_2^2 b_1^2 b_0^2 || b_5^1 b_4^1 b_3^1 b_2^1 b_1^1 b_0^1$.\smallskip

For $i,j \in \{0,1\}$, set $B_{(ij)} := \{s_5 s_4 s_3 s_2 s_1 s_0\}$ with $s_r = 1$ iff $b_r^2 = i$ and $b_r^1 = j$. That is, $B_{(ij)}$ is the characteristic vector of the column $(ij)^{T}$ in the 2 by $\frac{m}2$ array formed by $B_2$ and $B_1$ in parallel. Hence,

$$B_{(00)} = 010100,\ B_{(01)} = 000001,\ B_{(10)} = 001000,\ B_{(11)} = 100010.$$

\begin{figure}[!t]
\centerline{\includegraphics[scale=0.65,clip]{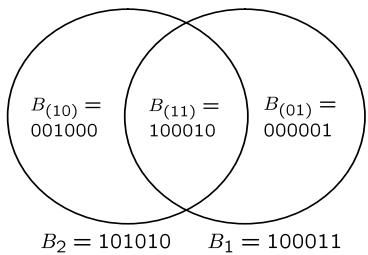}}
\caption{Venn diagram of characteristic vectors}
\label{fig1}
\end{figure}

Note that all of $B_{(00)}$, $B_{(01)}$, $B_{(10)}$, and $B_{(11)}$ are bitwise mutually exclusive, or disjoint. All these characteristic vectors except $B_{(00)}$ can be visualized in a natural way as a Venn diagram (see Fig. \ref{fig1}). Hence, $B_1$ and $B_2$ can be represented as

$$B_1 = \sum_{i\in\{0,1\}}\!\!\!B_{(i1)} = B_{(01)} + B_{(11)},\
  B_2 = \sum_{j\in\{0,1\}}\!\!\!B_{(1j)}= B_{(10)} + B_{(11)}.$$

Now, the multiplication of $A$ by $B$ can be parallelized essentially by multiplying $A$ by $B_{(01)}$, $B_{(10)}$, and $B_{(11)}$; the final assembly of the results of these multiplications requires a few additions and shifts. Namely,

\begin{displaymath}
\begin{split}
A \times B &= A \times (2^n \cdot B_2 + B_1) = 2^n (A \times B_2) + A \times B_1 \\
           &= 2^n (A \times B_{(10)} + A \times B_{(11)}) + A \times B_{(01)} + A \times
           B_{(11)},
\end{split}
\end{displaymath}

where $2^n \cdot z$ can be performed by an $n$-bit left shift of $z$.\smallskip

All these procedures are summarized in Algorithm 1. Note that Algorithm 1 eliminates the need of storage for characteristic vectors by combining the  partitioning into characteristic vectors and the parallel evaluation of several $A \times B_{(ij)}$ computations.

\begin{center}\label{alg1}
{Accumulate-and-add multiplication by operand-folding in
half}
\begin{tabbing}
123456\=1234\=1234\=1234\=1234\= \kill
{\bf Input:} $m$-bit integers $A$ and $B = B_2 || B_1$, where $B_i$ = ($b_{n-1}^i \cdots b_1^i b_0^i$) and $n = m/2$ \\
{\bf Output:} $C = A \times B$ \\
{\tt 1}\> $C_{(01)} \leftarrow C_{(10)} \leftarrow C_{(11)} \leftarrow 0$ \\
{\tt 2-1}\> {\bf for} $i=0$ {\bf to} $n-1$ {\bf do} \\
{\tt 2-2}\>\> {\bf if} ($b_i^2 b_i^1$) $\neq$ (00) \\
{\tt 2-3}\>\>\> $C_{(b_i^2 b_i^1)}$ $\leftarrow$ $C_{(b_i^2 b_i^1)} + A$ \\
{\tt 2-4}\>\> $A$ $\leftarrow$ $A \ll 1$ \\
{\tt 3-1}\> $C_{(10)}$ $\leftarrow$ $C_{(10)}$ + $C_{(11)}$\\
{\tt 3-2}\> $C_{(01)}$ $\leftarrow$ $C_{(01)}$ + $C_{(11)}$\\
{\tt 4}\> $C$ $\leftarrow$ ($C_{(10)} \ll n$) + $C_{(01)}$
\end{tabbing}
\end{center}

Suppose that both $A$ and $B$ are $m$-bit integers and each $B_i$ is
an $\frac{m}{2}$-bit integer. On average, the Hamming weights of
$B_i$ and $B_{(ij)}$ are $\frac{m}{4}$ and $\frac{m}{8}$,
respectively. For evaluating $A \times B$, Algorithm 1 requires
$\frac{3m}{8}+ 3$ additions without taking into account shift
operations into account. Hence, performance improvement over
classical accumulate-and-add multiplication is $\frac{m/2}{3m/8+3} \approx \frac{4}{3}$. In exchange,
Algorithm 1 requires three additional temporary variables.

\subsection{Generalized Operand Decomposition}

Let $B$ be an $m$-bit multiplier having the binary representation
$(b_{m-1} \cdots b_1 b_0)$, {\sl i.e.}, $B = \sum_{i=0}^{m-1}b_i2^i$ where
$b_i \in \{0,1\}$. By decomposing $B$ into $k$ parts, $B$ is split
into $k$ equal-sized substrings as $B = B_k || \cdots || B_2 ||
B_1$, where each $B_i$,
represented as $(b_{n-1}^i \cdots b_1^i b_0^i)$, is
$n = \lceil \frac{m}{k} \rceil$-bits long. If $m$ is not a multiple of $k$,
then $B_k$ is left-padded with zeros to form an $n$-bit string.
Hence,
\begin{equation}\label{eq_Horner}
A \times B = \sum_{i=1}^{k} 2^{n(i-1)}(A \times B_i).
\end{equation}

By Horner's rule, equation \ref{eq_Horner} can be rewritten as
\begin{equation}
A \times B = 2^n(2^n(\cdots(2^n(A \times B_k) + A \times B_{k-1})
\cdots ) + A \times B_2) + A \times B_1.
\end{equation}

The problem is now reduced into the effective evaluation of the $\{A
\times B_i \,|\, i=1,2,\ldots,k ;\, k \geq 2\}$ in advance, which is
known as the common-multiplicand multiplication ({\sc cmm}) problem. For
example \cite{YL93,WC95,KJ94} dealt with the case $k=2$,
and \cite{Yen97} dealt with the case $k=3$ or possibly
more. In this work we present a more general and efficient {\sc cmm}
method.

As in the toy example above, the first step is the generation of
$2^k$ disjoint characteristic vectors $B_{(i_k \cdots i_1)}$ from
the $k$ decomposed multipliers $B_i$. Each $B_{(i_k \cdots i_1)}$
is $n$ bits long and of average Hamming weight $n/2^k$.
Note that, as in Algorithm 1, no additional storage for the
characteristic vectors themselves is needed in the parallel computation of the $A
\times B_{(i_k \cdots i_1)}$'s.

The next step is the restoration of $A \times B_j$ for $1 \leq j
\leq k$ using the evaluated values $C_{(i_k \cdots i_1)} = A
\times B_{(i_k \cdots i_1)}$. The decremental combination method
proposed in \cite{Chung07} makes this step more efficient than other
methods used in {\sc cmm}. For notational convenience, $C_{(0 \cdots 0 i_j
\cdots i_1)}$ can simply be denoted as $C_{(i_j \cdots i_1)}$ by
omission of zero runs on its left side, and $C_{(i_k \cdots i_1)}$
can be denoted as $C_{(i)}$ where $(i_k \cdots i_1)$ is the binary
representation of a non-negative integer $i$. Then $A \times B_j$
for $j = k, \ldots, 1$ can be computed by
\begin{displaymath}
\begin{split}
&A \times B_j = \sum_{(i_{j-1}\cdots i_1)} C_{(1i_{j-1}\cdots i_1)},\\
&C_{(i_{j-1}\cdots i_1)} = C_{(i_{j-1}\cdots i_1)} +
C_{(1i_{j-1}\cdots i_1)},\quad \forall (i_{j-1} \cdots i_1).
\end{split}
\end{displaymath}

\begin{figure}[!t]
\centerline{\includegraphics[scale=0.6,clip]{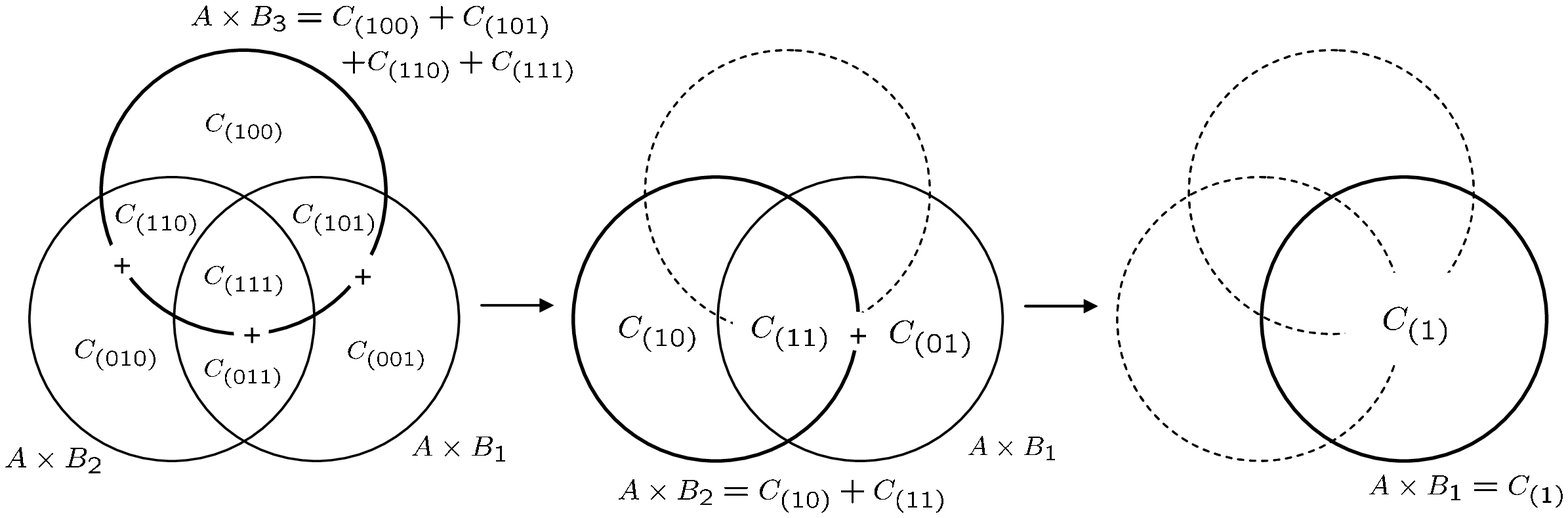}}
\caption{Venn diagram representation for combination process when
$k=3$} \label{fig2}
\end{figure}

Figure \ref{fig2} shows the combination process for a case $k=3$ with Venn diagrams.\smallskip

The last step is the application of Horner's rule on the results obtained from the above step. The overall procedure to compute $A
\times B$ is given in Algorithm 2. Note that Algorithm 2 saves memory by recycling space for evaluated characteristic vectors, without
use of temporary variables for $A \times B_i$.

\begin{center}\label{alg2}
{Accumulate-and-add multiplication by generalized operand
decomposition}
\begin{tabbing}
123456\=1234\=1234\=1234\=1234\= \kill {\bf Input:} $m$-bit integers
$A$ and $B = B_k || \cdots || B_1$, where $B_i$ = ($b_{n-1}^i \cdots
b_1^i b_0^i$) and $n = \lceil m/k \rceil$ \\
{\bf Output:} $C = A \times B$ \\
{\tt 1}\> $C_{(i_k \cdots i_1)} \leftarrow 0$ \quad for all $(i_k \cdots i_1) \neq (0 \cdots 0)$ \\
{\tt 2-1}\> {\bf for} $i=0$ {\bf to} $n-1$ {\bf do} \\
{\tt 2-2}\>\> {\bf if} ($b_i^k \cdots b_i^1$) $\neq$ $(0 \cdots 0)$ \\
{\tt 2-3}\>\>\> $C_{(b_i^k \cdots b_i^1)}$ $\leftarrow$ $C_{(b_i^k \cdots b_i^1)} + A$ \\
{\tt 2-4}\>\> $A$ $\leftarrow$ $A \ll 1$ \\
{\tt 3-1}\> {\bf for} $i=k$ {\bf down to} 1 {\bf do} \\
{\tt 3-2}\>\> {\bf for} $j=1$ {\bf to} $2^{i-1}-1$ {\bf do} \\
{\tt 3-3}\>\>\> $C_{(2^{i-1})}$ $\leftarrow$ $C_{(2^{i-1})} + C_{(2^{i-1} + j)}$ \quad \{$C_{(2^{i-1})}$ corresponds to $A \times B_i$\}\\
{\tt 3-4}\>\>\> $C_{(j)}$ $\leftarrow$ $C_{(j)} + C_{(2^{i-1} + j)}$ \\
{\tt 4-1}\> $C$ $\leftarrow$ $C_{(2^{k-1})}$ \\
{\tt 4-2}\> {\bf for} $i=k-1$ {\bf down to} 1 {\bf do} \\
{\tt 4-3}\>\> $C$ $\leftarrow$ $C \ll n$\\
{\tt 4-4}\>\> $C$ $\leftarrow$ $C + C_{(2^{i-1})}$
\end{tabbing}
\end{center}

\section{Theoretical Asymptotic Analysis}

It is interesting to determine how the actual number of additions necessary to perform a multiplication decreases as parallelization increases.
Neglecting the additions required to recombine the parallelized results, the number of additions tends to zero as the degree of parallelism $k$
increases. The convergence is slow, namely:

$$
\frac{\log k}k \sim \frac{\log\log m}{\log m}
$$

since $k<\log m$ is required to avoid edge effects. In practice if the operand is split into an exponential number of sub-blocks (actually $3^k$)
the total Hamming weight of the blocks will converge to zero.\smallskip

To understand why things are so, we introduce the following tools:\smallskip

Let $\delta_0\in[0,\frac12]$ and $\delta_{i+1}=\delta_i(1-\delta_i)$ then

$$\lim_{i\rightarrow\infty} \delta_i = 0$$

More precisely, $\delta_i = \theta(\frac{1}i)$ and

$$
\sum_{i=0}^{n-1} \delta_i^2=\delta_0-\delta_n
~~~\Rightarrow~~~
\sum_{i=0} \delta_i^2=\delta_0
$$

Let $B$ have length $b$ and density $\delta_i$, {\sl i.e.} weight $\delta_ib$. After performing the splitting process, we get three blocks,
$B_{(10)}$, $B_{(01)}$ and $B_{(11)}$ of length $\frac{b}{2}$ and respective densities $\delta_{i+1}=\delta_i(1-\delta_i)$ for the first
two and $\delta_i^2$ for $B_{(11)}$. The total cost of a multiplication is now reduced from $\delta_ib$ to

$$
\delta_ib-\frac{\delta_i^2b}{2}
$$

In other words, the gain of this basic operation is nothing but the Hamming weight of $B_{(11)}$:

$$
\frac{\delta_i^2b}{2}
$$

Graphically, the operation can be regarded as a tree with root $B$, two nodes $B_{(10)}$, $B_{(01)}$ and a leaf $B_{(11)}$. The gain is
the Hamming weight of the leaf.\smallskip

We will now show that by iterating this process an infinity of times, the total gain will converge to the Hamming weight of $B$.

\subsection{First Recursive Iteration of the Splitting Process}

Apply the splitting repeatedly to the nodes: this gives a binary tree having two nodes and one leaf at level one, and more generally
$2^j$ nodes and $2^{j-1}$ leaves at level $j$. The gain $\gamma_{1,j}$ of this process is the sum of the weights of the $N_{1,j}=2^j-1$
leaves, that is:

$$\frac{b}{2} \sum_{i=0}^{j-1}\delta_i=\frac{b}{2} (\delta_0-\delta_j)$$

As $j$ increases we get an infinite tree $A_1$, a gain of

$$\gamma_1=\frac{b\delta_0}{2}$$

and a total weight of

$$
W_1=b\delta_0-\frac{b\delta_0}2=\frac{b\delta_0}2
$$

\subsection{Second Recursive Iteration of the Splitting Process}

We now apply the previous recursive iteration {\sl simultaneously} (in parallel) to all leaves. Note that each leaf from the previous
step thereby gives rise to $1+2+\ldots+2^s+\ldots$ new leafs. In other words, neglecting edge effects we have $N_{2,j}\approx N_{1,j}^2$.\smallskip

The last step consists in iterating the splitting process $i$ times and letting $i$ tend to infinity. By analogy to the calculations of the
previous section the outcome is an extra gain of:

$$
\gamma_2=W_2=\frac{W_1}{2}
$$

Considering $W_t$ and letting $t\rightarrow\infty$, we get a total gain of:

$$
\Gamma=\sum_i \gamma_i = 2W_i=b\delta_0
$$

Thus a non-intuitive phenomenon occurs:

\begin{itemize}
\item Although $N_{i,j}\approx N_{1,j}^i$, eventually the complete ternary tree $T$ is covered, hence there are no pending leaves.\smallskip

\item The sum of an exponential number of weights ($3^k$ with $k\rightarrow\infty$) tends to zero.
\end{itemize}

\subsection{Speed of Convergence}

The influence of truncation to a level $k<\log n$ is twofold:

\begin{itemize}
\item The recursive iterations $R_i$ are limited to $i=k$, thus limiting the number of additional gains $\gamma_i$ to $\gamma_k$.\smallskip

\item Each splitting process is itself limited to level $k$, thus limiting each additional gain $\gamma_i, 1\leq i\leq k$ to $\gamma_{i,k}$.
\end{itemize}

Let us estimate these two effects:

$$
k<\log n-\log\log n \Rightarrow \Gamma_k = \sum_{i=1}^k < \delta_0 (1-\frac{\log n}{n})
$$

$$
k>\log n-\log\log n \Rightarrow  = \sum_{i=1}^k \gamma_i-\gamma_{i,k} >(\log n-\log\log n) \min(\gamma_i-\gamma_{i,k})
$$

But $$\min(\gamma_i-\gamma_{i,k})\approx \frac{1}{2n}(1-o(1))$$

Hence the global weight tends to zero like $\theta(\frac{\log k}{k})$.

\section{Performance Analysis and Comparison}

Accumulate-and-add multiplication performance is proportional
to the number of additions required. Hence, we analyze the
performance of the proposed multiplication algorithm.\smallskip

In step 2, as the average Hamming weight of each characteristic
vector is $n/2^k$, where $n = \lceil m/k \rceil$, the number of
additions needed to multiply $A$ by $2^k-1$ disjoint characteristic
vectors in parallel is $(2^k-1)\cdot \frac{n}{2^k}$ on average. In
step 3, the computation of every $A \times B_i$ by combination of
the evaluated characteristic vectors requires the following number
of additions:
$$\sum_{i=1}^{k} 2(2^{i-1}-1) = \sum_{i=1}^{k}(2^i-2) = 2^{k+1} -
2k -2,$$

whereas the method used in \cite{Yen97} requires
$k(2^{k-1}-1)$ additions. In step 4, the completion of $A \times B$
using Horner's rule requires $k-1$ additions. Therefore, the
total number of additions needed to perform the proposed algorithm
is on average equal to:
$$F_{\mbox{{\scriptsize avg}}}(m,k) = \frac{2^k-1}{2^k}\cdot \left\lceil \frac{m}{k}
\right\rceil + 2^{k+1} - k - 3.$$ On the other hand, $F_{\mbox{{\scriptsize wst}}}(m,k) =
\left\lceil \frac{m}{k} \right\rceil + 2^{k+1} - k - 3$ in the worst
case.\smallskip

\begin{table}[!t]
\caption{Optimal $k$ for $F$ as a function of $m$} \label{tab1}
\begin{center}
\begin{tabular}{|c||c|c|c|}
    \hline
    ~~Optimal $k$~~& Range of $m$~~&~~$\frac{m}2 F_{\mbox{{\scriptsize avg}}}(m,k)/F_{\mbox{{\scriptsize avg}}}(m,1)$~~&~~$mF_{\mbox{{\scriptsize wst}}}(m,k)/F_{\mbox{{\scriptsize wst}}}(m,1)$~~\\ \hline \hline
    2             & $24 \leq m \leq 83$     & $0.375m+3$                                                                      & $0.500m+3$  \\ \hline
    3             & $84 \leq m \leq 261$    & $0.292m+10$                                                                     & $0.333m+10$ \\ \hline
    4             & $262 \leq m \leq 763$   & $0.234m+25$                                                                     & $0.250m+25$ \\ \hline
    5             &~~$764 \leq m \leq 2122$~~& $0.194m+56$                                                                     & $0.200m+56$ \\ \hline
\end{tabular}
\end{center}
\end{table}

Performance improvement over the classical accumulate-and-add
multiplication algorithm is asymptotically:

$$ \lim_{m \to \infty} \frac{F_{\mbox{{\scriptsize avg}}}(m,1)}{F_{\mbox{{\scriptsize avg}}}(m,k)} = \lim_{m \to \infty} \frac{m/2}{\frac{2^k-1}{2^k}\cdot \left\lceil \frac{m}{k}
\right\rceil + 2^{k+1} - k - 3} = \frac{k \cdot 2^{k-1}}{2^k - 1}.$$

Larger $k$ values do not necessarily guarantee the better
performance, because the term $2^{k+1}-k-3$ increases exponentially
with $k$. Thus, a careful choice of $k$ is required. The
analysis of $F_{\mbox{{\scriptsize avg}}}$ for usual multiplier sizes $m$ yields optimal $k$ values
that minimize $F_{\mbox{{\scriptsize avg}}}$. The optimal $k$ values as a function of $m$ are given in Table \ref{tab1}.
Table \ref{tab1} also includes comparisons with the classical algorithm for the both the case and the worst cases.\smallskip

In modern public key cryptosystems, $m$ is commonly chosen between 1024 and 2048. This corresponds to the optimum $k=5$ {\sl i.e.} an
2.011 to 2.260 performance improvement over the classical algorithm and 1.340 to 1.560 improvement over the canonical signed digit
multiplication algorithm \cite{AW93} where the minimal Hamming weight of is $\frac{m}3$ on the average.\smallskip

On the other hand, the proposed algorithm requires storing $2^k - 1$ temporary variables, which correspond to
$O((2^k-1)(m+n+k))$-bit memory. Whenever $k \geq 3$, although optimal performance is not guaranteed, the new algorithm is still
faster than both classical and canonical multiplication.

\newpage

\appendix
\section{Hardware Implementation}

{\scriptsize
\begin{verbatim}

LIBRARY IEEE; USE ieee.std_logic_1164.all; USE ieee.std_logic_unsigned.all;

ENTITY Mult_Entity IS
    GENERIC(CONSTANT m : NATURAL := 32;
            CONSTANT k  : NATURAL := 2);
    PORT(A : in STD_LOGIC_VECTOR (m-1 DOWNTO 0);
         B : in STD_LOGIC_VECTOR (m-1 DOWNTO 0);
        C : out STD_LOGIC_VECTOR(2*m-1 DOWNTO 0));

END Mult_Entity;
ARCHITECTURE Behavioral OF Mult_Entity IS
    SIGNAL n : NATURAL := m+k-1/k;
    SIGNAL INPUT_LENGTH : NATURAL := n*k;
    SIGNAL OUTPUT_LENGTH : NATURAL := 2*INPUT_LENGTH;
    SIGNAL C_TEMP : STD_LOGIC_VECTOR(2*INPUT_LENGTH-1 DOWNTO 0);
    SIGNAL C_PARTS_LENGTH : NATURAL := INPUT_LENGTH+n;
    SIGNAL A_TEMP : STD_LOGIC_VECTOR(C_PARTS_LENGTH-1 DOWNTO 0);
    SIGNAL B_value : INTEGER;
    TYPE BX_TYPE IS ARRAY (k DOWNTO 1) OF STD_LOGIC_VECTOR(n-1 DOWNTO 0);
    SIGNAL BX : BX_TYPE;
    SIGNAL cx_count : NATURAL := 2**k-1;
    TYPE CX_TYPE IS ARRAY (cx_count DOWNTO 1) OF STD_LOGIC_VECTOR(C_PARTS_LENGTH-1 DOWNTO 0);
    SIGNAL CX : CX_TYPE;

BEGIN
Myproc : PROCESS(A,B)
    VARIABLE i, j : INTEGER := 0;
BEGIN

FOR i IN 1 TO k-1 LOOP BX(i)(n-1 DOWNTO 0) <= B(i*n-1 DOWNTO (i-1)*n); END LOOP;
    BX(k)(m-(n*(k-1))-1 DOWNTO 0) <= B(m-1 DOWNTO m-n*(k-1));
IF ((m MOD k)>0) THEN BX(k)((n-1) DOWNTO (n-1-(m MOD k))) <= "0"; END IF;
A_TEMP (m-1 DOWNTO 0) <= A; A_TEMP (C_PARTS_LENGTH-1 DOWNTO m) <= "0";

--STEP 1
For i IN 1 TO 2**k-1 LOOP CX(i) <= "0"; END LOOP;
--STEP 2-1
For i IN 0 TO n-1 LOOP
    B_value <= 0;
    FOR j IN 1 TO k LOOP
        IF ((BX(j)(i))='1') THEN B_value <= B_value + 2**(j-1); END IF;
    END LOOP;
--STEPS 2-2 and 2-3
    IF (B_value>0) THEN CX (B_value) <= CX (B_value) + A_TEMP; END IF;
--STEP 2-4
    A_TEMP <= A_TEMP(C_PARTS_LENGTH-2 DOWNTO 0)&"0";
END LOOP;

--STEP 3-1
FOR i IN k DOWNTO 1 LOOP
--STEP 3-2
  FOR j IN 1 TO 2**(i-1)-1 LOOP
--STEP 3-3
    CX(2**(i-1)) <= (CX(2**(i-1)) + CX(2**(i-1)+j));
--STEP 3-4
    CX(j) <= (CX(j) + CX(2**(i-1)+j));
  END LOOP;
END LOOP;
--STEP 4-1
C_TEMP (C_PARTS_LENGTH-1 DOWNTO 0) <= CX(2**(k-1));
C_TEMP (n-1 DOWNTO C_PARTS_LENGTH-1) <= "0" ;
--STEP 4-2
FOR i IN k-1 DOWNTO 1 LOOP
--STEP 4-3
    C_TEMP <= C_TEMP(2*m-1-n DOWNTO 0) & "0" ;
--STEP 4-4
    C_TEMP <= C_TEMP + CX(2**(i-1));
END LOOP;

END PROCESS Myproc;
C <= C_TEMP;
END Behavioral;
\end{verbatim}}


\begin{thebibliography}{100}

\bibitem{YL93} Yen, S.-M., and Laih, C.-S., Common-multiplicand multiplication and its applications to public key cryptography, {\em Electron. Lett.}, 1993, 29(17), pp. 1583--1584.

\bibitem{WC95} Wu, T.-C., and Chang, Y.-S., Improved generalisation common-multiplicand multiplications algorithm of Yen and Laih, {\em Electron. Lett.}, 1995, 31(20), pp. 1738--1739.

\bibitem{Yen97} S.~Yen, Improved common-multiplicand multiplication and fast exponentiation by exponent decomposition, {\em IEICE Trans. Fundamentals}, 1997, {\sc E80-A}(6), pp. 1160--1163.

\bibitem{KJ94} C.~Ko\c{c} and S.~Johnson, Multiplication of signed-digit numbers, {\em Electron. Lett.}, 1994, 30(11), pp. 840--841.

\bibitem{LC96} D.~Lou, and C.~Chang, Fast exponentiation method obtained by folding the exponent in half, {\em Electron. Lett.}, 1996, 32(11), pp. 984--985.

\bibitem{Chung07} Chung, B., Hur, J., Kim, H., Hong, S.-M., and Yoon, H., Improved Batch Exponentiation, {\em Submitted to Inform. Process. Lett}, Nov. 2005.\\
{\scriptsize {\tt http://nslab.kaist.ac.kr/$\sim$bcchung/publications/batch.pdf}}

\bibitem{AW93} Arno, S., and Wheeler, F.S., Signed digit repersentations of minimal Hamming weight, {\em IEEE Trans. Computers}, 1993, 42(8), pp. 1007--1010.
\end{thebibliography}
\end{document}